# 5G for Railways: the Next Generation Railway Dedicated Communications

Ruisi He[1], Bo Ai[1], Zhangdui Zhong[1], Mi Yang[1], Ruifeng Chen[2], Jianwen Ding[1],
Zhangfeng Ma[1], Guiqi Sun[1], and Changzhu Liu[1]
[1] State Key Laboratory of Rail Traffic Control and Safety, Beijing Jiaotong University, China
[2] Institute of Computing Technology, China Academy of Railway Sciences, China

*Abstract*—To overcome increasing traffic, provide various new services, further ensure safety and security, significantly improve travel comfort, a new communication system for railways is required. Since 2019, public networks have been evolving to the fifth generation communication (5G) worldwide, whereas the main communication system of railway is still based on the second generation communication (2G). It is thus necessary for railways to replace the current 2G-based technology with the next generation railway dedicated communication system with improved capacity and capability, and the 5G for railways (5G-R) technology is a promising solution for further intelligent railways. This article gives a review of the current developments of the next generation railway communications, followed by a discussion of the typical services that the 5G-R can provide to intelligent railways. Then, main application scenarios of 5G-R are summarized and system configurations are compared. Some key technologies of 5G-R such as network architecture, massive MIMO, millimeter-wave, multiple access scheme, ultra-reliable low latency communication, and advanced video processing are presented and analyzed. Finally, some challenges of 5G-R are highlighted.

## I. INTRODUCTION

Railway has been an important mode of transportation during the development of human society. Tracing back to the first industrial revolution, human industrial technology has experienced mechanical revolution, electric power revolution, and information technology revolution. Nowadays, we are going through intelligent technology revolution, owing to the fast developments of artificial intelligence, big data, cloud computing, and the fifth generation communication (5G). Meanwhile, railway is about to move from information era to intelligent era. Intelligent railway will apply 5G, artificial intelligence, big data, cloud computing, Internet of Things (IoTs), satellite navigation and other new-generation information technologies. Through comprehensive and efficient use of resources, intelligent railways can realize comprehensive sensing, ubiquitous interconnection, and fusion processing of railway equipment, infrastructure, and environmental information. Railway system will enter a period of safer, more efficient, greener, more comfortable, and faster development.

The International Union of Railways (UIC) has been advocating railway digital and intelligent transformation and developing a new platform. The Future Railway Mobile Communication System (FRMCS) [1] is proposed by UIC, as a key enabler for railway digitalization. UIC further releases the FRMCS User Requirements Specification (URS) [2], which includes 72 use cases for the further railway communications. Such huge amounts of new services and business needs cannot be supported by the current Global System for Mobile Communications - Railways (GSM-R). After UIC released FRMCS, the 3rd Generation Partnership Project (3GPP) made a study to determine the working requirements of FRMCS and analyze the differences between it and the existing functions in 3GPP [3]. Since 3GPP Release 15 only includes Enhanced Mobile Broadband, most of the requirements of FRMCS can only be well considered and implemented in Release 16 and Release 17. Therefore, intelligent railway communication system should be based on Release 16 and has the ability to smoothly evolve to Release 17.

Meanwhile, many countries have actively formulated their own development strategies of intelligent railways to enhance integration and innovative application of intelligent technologies and railways. The European Union (EU) proposed the 5GRAIL project [4], which is organized by UIC and other companies such as the European Rail Industry, Nokia, Siemens, and Alstom. 5GRAIL will establish partnership with EU Shift2Rail to meet changing EU transport needs, by using advanced and intelligent technologies. Germany proposed Industry 4.0 and Strong Rail strategy, which copes with massive railway technological changes, such as intelligent railway production, operation, and maintenance. Switzerland proposed SmartRail 4.0 to modernize railway intelligent system with the aim of ensuring robustness of future services, increasing capacity of the existing infrastructure, and increasing safety of employees in track area. The Cyber-Rail proposed by Japan is committed to providing real-time passenger information service and making train operation more flexible, safer, more reliable and more competitive. NTT DOCOMO and East Japan Railway Company have verified stable operation of typical 5G communication capabilities, including handover between base stations and transmission of high-definition video data. Korea Railroad Research Institute has tested 5G-based autonomous train control technology. It has been a common consensus among countries to apply digital technology to the future railway systems and make better use of the advantages of intelligent railways.

Nowadays, the existing GSM-R, as a narrow-band communication system with low bandwidth and data rate, cannot meet the requirements of smart railway services such as multimedia dispatching communication, intelligent operation and maintenance, etc [5]. GSM-R is also facing the problem of insufficient industrial chain support, and it urgently needs to evolve to a new generation of railway communication system [6]. LTE-Railways (LTE-R) had been considered for railway communication system evolution at the beginning of 2010s [7], however, it is found that the LTE system cannot meet the requirements of ultra-reliable communication and massive connectivity for future railways. It also cannot support the smart services and intelligent applications in future railways [6]. Meanwhile, the fast development of 5G all over the world will affect LTE product life cycle and supply chain, and bring more uncertainty, while it is important for railway industry to have stable supply chain and a long product life cycle. To satisfy high data rate and massive connectivity requirements, 5G for railways (5G-R) has attracted much attention nowadays. The high-data-rate, low-latency, and high-access-density characteristics of 5G can provide high quality services for intelligent railway applications. In Europe, the GSM-R system is obsolete and will be replaced by FRMCS, and the developments are ongoing. FRMCS will be bearer flexible and will revolutionize railway communications, and 5G can be a critical enabler for it to take the operations into the digital age [8]. China has issued a strategy to strengthen national transportation, and promote the digital and intelligent railway development based on 5G technologies [9]. At present, China's railway system is moving towards the stage of intelligent operation, and gradually forming China's intelligent railway system.

In summary, 5G-R is a promising solution for further intelligent railways, and this article aims to review and analyze the features and key challenges of 5G-R. The rest of the article is organized as follows. The next section presents the typical services that 5G-R can provide to intelligent railways. Then, the main application scenarios of 5G-R are summarized. Following that we discuss some key technologies of 5G-R. Finally, some challenges of 5G-R are highlighted.

## II. SERVICES OF 5G-R

The existing railway dedicated communication system such as GSM-R only provides data transmission for control signals, and mainly covers train-to-ground communication environments. The services of 5G-R are diverse, and the future railway communications should provide reliable wireless coverage in more cases including continuous wide-area coverage along railway lines, railway yards and hot spots coverage, monitoring of railway ground infrastructure, and broadband smart applications for intelligent trains. The services of 5G-R based intelligent railways can be generally divided into four categories, which are presented in detail as follows.

(1) *Train Operation Related Service*: It includes multimedia scheduling, railway dispatching command, automatic train operation, locomotive synchronous control, train number information checking, warning of train approaching, shunting train protection, etc. Such services are especially important to the safety of train operation.

(2) *Maintenance Service*: It includes emergency communications, intelligent maintenance communications, data update of on-board equipment, railway freight information transmission, passenger services, etc. These services are helpful to improve railway intelligent operation and maintenance.

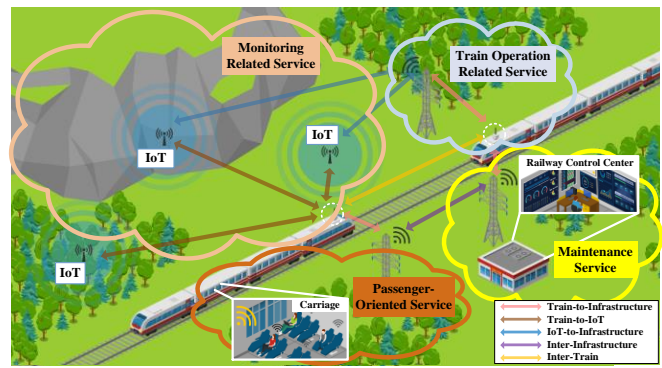

**Fig. 1**. Fully connected 5G-R network architecture and services.

(3) *Monitoring Related Service*: It includes real-time monitoring of train status and on-board equipment, power safety monitoring, locomotive driver monitoring, overhead power line monitoring, construction vehicle monitoring, railway cables monitoring, etc. Such services can improve safety of railway system.

(4) *Passenger-Oriented Service*: It includes high-speed Internet on train and on platform, real-time and intelligent passenger information system, multimedia personal infotainment (i.e., information and entertainment) in railway environments, visual/audio public address information, etc. Such services can improve passenger comfort and travel efficiency.

Various services and wireless coverage environments put forward higher requirements on the bandwidth, delay, reliability, and security of communication system. They also put forward new requirements for information sharing, and a unified information and communication platform is thus needed to connect various devices in railway environments. 5G-R should support such comprehensive services, and provide complete coverage and wireless access in railway lines, tunnels, stations, railway yards and depots, carriages, etc. Fig. 1 shows a fully connected 5G-R network architecture and the corresponding services, where comprehensive connectivity is supported including train-to-ground link, train-to-train link, and IoT, for all the railway-related devices. In addition, it is challenging to offer all the services with limited bands and low deployment cost for 5G-R.

## III. APPLICATION SCENARIOS OF 5G-R

5G-R motivates new applications in future intelligent railways by providing rich wireless services and improved coverage. Application scenarios of 5G-R can be generally divided into four categories: intelligent construction, intelligent equipment, intelligent operation and maintenance, and smart travel. Each application scenario and the corresponding information and communication requirements are shown in Table I. The situation of the supported communication systems are also compared in Table I, and it is found that only 5G-R can support all the intelligent application scenarios in future railway systems.

Table I TYPICAL 5G-R APPLICATION SCENARIOS

| Application Scenario | Technical Requirement & System Support | Technical Requirements | | | | | | | System Support | | |
|---|---|---|---|---|---|---|---|---|---|---|---|
| | | Audio and video reliable transmission | IoT data collection | IoT real-time control | Precise positioning | High speed Audio and video stream | Offloading with large bandwidth | Augmented reality | GSM-R | LTE-R | 5G-R |
| **Intelligent Construction** | Intelligent survey and design | | | | √ | | | √ | | | √ |
| | Dispatch communication | √ | | | √ | | | | √ | √ | √ |
| **Intelligent Equipment** | Intelligent multiple-unit train | | √ | √ | √ | | √ | | | | √ |
| | Train control system | √ | | | √ | | | | √ | √ | √ |
| **Intelligent Operation and Maintenance** | On-board equipment monitoring | √ | √ | √ | √ | | √ | | | √ | √ |
| | Ground infrastructure monitoring | | √ | √ | √ | | | | | √ | √ |
| | Multimedia dispatch communications | √ | | | √ | √ | | √ | | √ | √ |
| | Emergency communications | √ | | | √ | | | | √ | √ | √ |
| | Safety-related perception and protection | | √ | √ | √ | | | | | | √ |
| | Intelligent video surveillance | | | | √ | √ | | | | | √ |
| **Smart Travel** | Multimedia entertainment | √ | | | | √ | √ | √ | | | √ |
| | Station navigation | | √ | | √ | | | | | | √ |
| | Smart logistic | √ | √ | √ | √ | | | | | | √ |

(1) *Intelligent Construction*: It unifies railway design and construction stages and adopts new technologies such as building information modeling (BIM), digital design and management, IoT environment monitoring, high-precision staff and equipment positioning, etc. Intelligent construction can improve the efficiency of railway architectural design and construction, realize sustainable development of railway construction industry, and efficiently unify the various stages of architectural design, building component production, and on-site construction. Such applications require high-precision positioning, low end-to-end delay, low packet loss rate, large bandwidth and high data rate. This application scenario has been adopted by the development strategies of EU 5GRAIL, Korea Railroad BIM 2030, etc.

(2) *Intelligent Equipment*: It refers to railway equipment based on advanced technologies, such as intelligent technology, sensor technology, automation technology, and information technology. Intelligent equipment has the characteristics of ubiquitous perception, intelligent fusion, and analysis and decision making. Intelligent equipment in 5G-R mainly includes intelligent multiple-unit train and the next generation train control system. The former involves on board IoT, security and surveillance system, entertainment, passenger information system, etc. The latter requires a 5G network to provide high reliable communications with low latency and high data rate. This application scenario has been adopted by the development strategies of EU 5GRAIL, Switzerland SmartRail 4.0, etc.

(3) *Intelligent Operation and Maintenance*: It mainly includes railway intelligent maintenance, multimedia dispatch communication and emergency command. 5G-R will use low-power wireless sensors and wireless charging technology to lighten trackside equipment and realize green monitoring along railway lines. For railway intelligent maintenance, intelligent analysis of equipment status data and historical information, and on-site augmented reality assistance will be employed to realize the prediction of equipment failures, and further realize visualized digital twin operation and maintenance of railways. Moreover, 5G-R will support more intelligent and convenient multimedia dispatch communication and emergency command, and realize real-time audio and video dispatch command, and safety control of all elements in railway operation process [6]. 5G-R system needs to provide large connection and wide coverage in railway environments to support the large-scale sensing and interconnection required by the related application scenarios. This application scenario has been adopted by the development strategies of EU 5GRAIL, Germany Industry 4.0, Switzerland SmartRail 4.0, etc.

(4) *Smart Travel*: In the future, 5G-R will provide widely-covered communications with high data rate for the scenarios of railway station service and train passenger service, to realize smart travel. 5G-R supports the building of the smart railway station to better meet the needs of passengers in station for multimedia communication, station navigation, and smart feeder transportation. In the future smart stations, passengers entering the station will be more intelligent and efficient. Face recognition technology, smart security for passenger and baggage screening, and smart ticketing will significantly improve transportation efficiency. 5G-based high-precision positioning and navigation in station, intelligent monitoring of passenger flow, and robot guidance service will improve passenger service level and travel experience in railway station. Moreover, on-board passengers will enjoy the rich services based on 5G-R including multimedia entertainment, high-definition video broadcasting, virtual reality, mobile edge computing, etc., and the travel experience will be significantly improved. In addition, the capabilities of railway logistics real-time tracking and resource allocation will also be significantly enhanced by 5G-R. This application scenario has been adopted by the development strategies of EU 5GRAIL, Germany Industry 4.0, Japan Cyber-Rail, etc.

Table II SYSTEM CONFIGURATIONS OF GSM-R, LTE-R, 5G, AND 5G-R.

| System Parameter | GSM-R | LTE-R | 5G | 5G-R |
|---|---|---|---|---|
| Frequency | Uplink: 876-880 MHz Downlink: 921-925MHz | 450 MHz, 800 MHz, 1.4 GHz, 1.8 GHz | Sub-6GHz: 410-7125 MHz Millimeter-wave: 24.25-71 GHz | 900 MHz, 1.9 GHz, 2.1 GHz |
| Bandwidth | 0.2 MHz | 1.4-10 MHz | Sub-6GHz: 5-100 MHz Millimetter-wave: 50-2000 MHz | 10-20 MHz; possible ≥100 MHz at higher frequency bands |
| Modulation | GMSK | QPSK, 16-QAM | 256-QAM, OFDM, FBMC, UFMC, GFDM, f-OFDM | QPSK, 16-QAM, 64-QAM, 256-QAM, OFDM |
| Typical Cell Range | 8 km | 4-12 km | tens of meters to several kilometers | 1-6 km |
| MIMO | No | 2×2 | 16×16, 32×32, 64×64, 128×128 | 4×4, 8×8, 16×16, 32×32 |
| Average Data Rate | <10 Kbps | 1-10 Mbps | 50-100 Mbps | 10-50 Mbps |
| Peak Data Rate | 172 Kbps | Uplink: 10 Mbps Downlink: 50 Mbps | Uplink: 10 Gbps Downlink: 20 Gbps | Uplink: 50 Mbps Downlink: 200 Mbps |
| Peak Spectral Efficiency | 0.33 bps/Hz | 2.55 bps/Hz | Uplink: 15 bps/Hz Downlink: 30 bps/Hz | Uplink: 8 bps/Hz Downlink: 15 bps/Hz |
| Mobility | Max. 500 km/h | Max. 500 km/h | Max. 500 km/h | Max. 500 km/h |
| Reliability | 99.999% | 99.999% | 99.999% | 99.9999% |
| Connection Loss Rate | $\leq 10^{-2}$/h | $\leq 10^{-2}$/h | NA | Voice Related: $\leq 10^{-2}$/h Video Related: $\leq 10^{-3}$/h |
| End to End Latency | 500 ms | 200-500 ms | 1-10 ms | 50-500 ms |
| Handover Success Rate | ≥ 99.5% | ≥ 99.5% | 90- 99.5% | ≥ 99.9% |
| 4K Video | No | No | Support | Possible Support with Increased Bandwidth |

## IV. KEY TECHNOLOGIES OF 5G-R

The future railway system will develop in the direction of networking, intelligence, and automation. The next generation railway dedicated communication system should realize comprehensive perception, interconnection, and information interaction among all railway users and infrastructures. Typical intelligent railway applications, such as video-based track monitoring, ultra-high reliability train control, and intensive access of massive users and sensors, correspond to the three typical 5G scenarios respectively: Enhanced Mobile Broadband, Ultra-Reliable Low Latency Communication, and Massive Machine-Type Communication. Therefore, the development of railway dedicated communications supported by 5G has been widely expected. Table II summarizes some key parameters of GSM-R, LTE-R, 5G and 5G-R, and it is found that 5G-R has improved performance compared with the existing railway communication systems. The 5G-R system is expected to meet diversified and stringent railway communication requirements under the support of various key technologies.

(1) *Novel Network Architecture and Network Slicing*: Considering the complex and diverse railway applications, 5G-R can introduce a cell-free wireless access topology to provide more flexible wireless control, service monitoring, and protocol stack customization capabilities. Simultaneously, the novel network architecture can provide highly customizable network services for different users and applications, and establish a resource-sharing integrated information platform. The 5G-R network can implement multiple network slices on the same physical infrastructure platform and provide comprehensive slice division for applications with different requirements such as high-definition video surveillance, automatic train operation, and railway IoT, and finally realize the end-to-end slicing deployment of the railway dedicated communication network. It is noteworthy that 5G-R should guarantee safety services cohabitation with other services if public 5G network is used [6]. Another approach is to use 5G private network with possible share of infrastructure or hybridization with other safety applications.

(2) *Massive MIMO*: Centralized massive MIMO can generate narrow beams with large gain to high-speed trains and greatly improves system capacity. Moreover, distributed massive MIMO technology is expected to be a promising solution to deal with communication quality degradation problem caused by users' rapid movement and frequent handover. A distributed massive MIMO system deploys access points along the railway and transmits data to the central processing unit through fronthaul links. The advantages of the distributed massive MIMO include: i) Short distance between users and access points can reduce the influence of large-scale channel fading and provides reliable user Quality of Service (QoS). ii) Low-cost access points are suitable for flexible deployment. iii) Frequent handover can be avoided, thereby saving wireless resources. Fig. 2 shows average spectral efficiency of a recent proposed cell-free massive MIMO-OFDM system [10] for high-speed railway communications, where 8 distributed antennas are placed on top of different cars of a train, 20 distributed access points are placed along rail track and each access point is equipped with *N* antennas. It is found that the distributed massive MIMO can achieve fairly high spectral efficiency in high mobility scenario and the spectral efficiency decreases with moving speed due to the Doppler effect. Increasing the number of antennas per railway access point can also improve spectral efficiency.

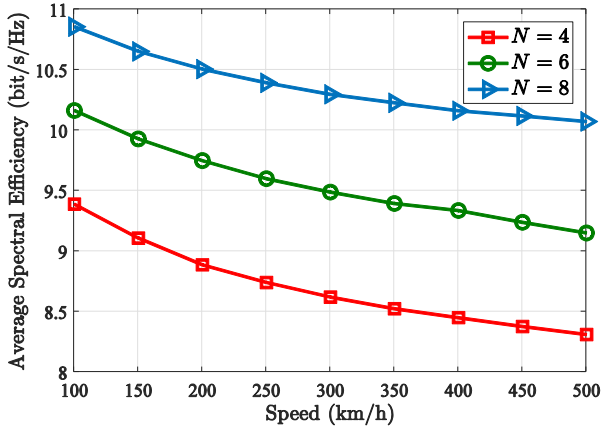

**Fig. 2**. Simulations of average spectral efficiency against different speeds of train in cell-free massive MIMO-OFDM systems. Carrier frequency is 2.1 GHz, bandwidth is 20 MHz, transmitted power is 200 mW, and Rice $K$-factor is 30 dB.

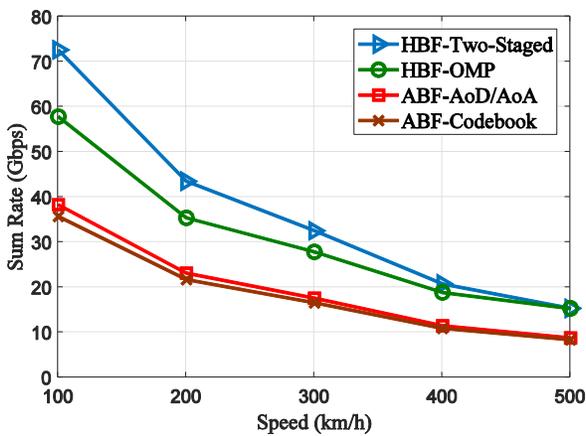

**Fig. 3**. Simulations of system sum rate against different speeds of train. Carrier frequency is 28 GHz and bandwidth is 500 MHz. HBF-Two-Staged is the two-stage hybrid beamforming algorithm proposed in [12], OMP represents orthogonal matching pursuit method, ABF-AoD/AoA represents analog beamforming scheme of tracking the angle of departure/arrival of the dominant paths, and ABF-Codebook represents the Codebook-based beam switching.

(3) *Millimeter-Wave*: Millimeter-wave technology can well use abundant bandwidth resource and support transmission data rate up to Gbps level for massive 5G-R applications. To combat the huge penetration loss and propagation attenuation in millimeter-wave bands, beamforming technique is utilized thereby realizing high-gain millimeter-wave transmission, and the repeater-nodes at the top of train can also be used to improve signal transmission [11]. The key challenges faced with its practical implementation mainly include the rapidly time-varying channel in high-mobility scenarios and random blockage events in complicated scenarios. With the known route and measurable position of trains, dynamic beam tracking is guided by prediction and hybrid beamforming is efficiently developed. It is shown in Fig. 3 that efficient beamforming can achieve tens of Gbps level sum data rate in train-to-ground millimeter-wave communications, and data rate generally decreases with moving speed due to the Doppler effect [12]. Considering realistic deployments such as the large-scale antenna system and the non-stationary channel condition, efficient beamforming design deserves more attention for single- and multiple-user scenarios.

(4) *Railway IoT*: In the near future, railway IoT will be equipped with massive sensors to enable comprehensive situation awareness. Generally, the massive IoT devices are with low power and low cost so that grant-free random access can be used to reduce signaling overhead. Also, due to the scarce radio resource, reliable grant-free random access for massive users cannot be supported with conventional multiple access schemes. Thus, non-orthogonal multiple access (NOMA) is used to achieve larger user access number. Among the NOMA schemes, tandem spreading multiple access (TSMA) can effectively realize reliable grant-free random access for massive railway IoT user devices [13].

(5) *Ultra-Reliable Low Latency Communication*: The future railway needs to guarantee less than 100 ms end-to-end latency for operation related voice and data communications, and support 99.9999% reliability for critical data communication, both at 500 km/h [14]. Ultra-reliable low latency communication mainly guarantees railway communication performance from two aspects: i) High reliable transmission under incomplete channel state information. Considering the impact of channel state information error, the transmission delay and packet error rate can be reduced by increasing transmission power and balancing inter-cell interference to ensure end-to-end transmission performance. ii) High reliable transmission with hybrid automatic repeat request. Considering the spectrum efficiency and energy efficiency of 5G-R system, as well as the small-packet feature of ultra-reliable low latency services, the finite block length coding theory can be adopted to characterize the transmission performance, and thus the system can be optimized with a unified framework to guarantee the requirements of railways.

(6) *Advanced Video Processing*: Comprehensive and intelligent video analysis and processing will significantly improve accuracy of detection and monitoring of railway environments and equipment, and promote the safe and intelligent developments of railways. In 5G-R, advanced video processing technology, e.g., deep learning based dehazing, dark light enhancement, noise reduction, deblurring, super-resolution restoration and segmentation, will be applied to railway perimeter intrusion detection, equipment and environment intelligent monitoring, key component status detection of railway track, pantograph status analysis, etc. In addition, intelligent video processing and data mining can also be used for railway security and counter terrorist, smart security for passenger and baggage screening, and passenger flow analysis and prediction.

## V. CHALLENGES

### A. Frequency Resource

Future railway communications should provide reliable data transmission for different applications such as train control, station and yard coverage, ground infrastructure monitoring, on-board broadband access, etc. Different communication services and application scenarios have different requirements of carrier frequency and bandwidth. Nowadays, many countries have not even allocated 5G dedicated frequency bands yet, and operators are promoting 5G-R public-private network convergence solutions. At the same time, railway industry is still actively striving for the private frequency of 5G-R networks. It is expected that the bandwidth that can be allocated to 5G-R will be smaller than that of the 5G public networks, which is an important factor affecting the performance of 5G-R. It is thus challenging to support the various applications and services of 5G-R with reduced frequency resource. In China, the potential 5G-R frequency band for train control is 1.9/2.1 GHz with 10 MHz bandwidth. Some millimeter-wave bands such as 28 GHz, 39 GHz, and 71-86 GHz are considered for enhanced broadband access and video transmission. UIC has also led efforts to obtain dedicated frequencies for 5G in Europe railway within the framework of FRMCS programme, and the Electronic Communications Committee approved the draft recommendation allocating 5.6 MHz in 900 MHz and 10 MHz in 1900 MHz for European railways.

*B. Reliable Big Data Transmission*

High speed train generally includes many users and devices, and instantaneous big data transmission with high reliability leads to significant challenges to 5G-R due to the following reasons.

*Non-Stationary Channel*: The channel of high-speed railway has been generally considered to be non-stationary due to the dynamic evolution of multipath components caused by motion. Non-stationary channel characterization has many advantages such as improving accuracy of channel simulation and system evaluation, pre-conditioning the waveform to match the expected fading profile, etc., which are important to guarantee reliable big data transmission. The dynamic behaviors should be well included in high-speed railway channel modeling and channel estimation.

*Carriage Penetration Loss*: The average carriage penetration loss of a typical high-speed train is about 25-35 dB for sub-6 GHz. Such high penetration loss significantly reduced SNR for on-board passengers, and the performance of on-board broadband service is thus affected. The condition may be even worse if millimeter-wave is used for data transmission. The penetration loss is also affected by the incidence angle of signal from base station to train. The link budget of 5G-R should be carefully conducted. A possible solution is to use mobile femtocells technology to enhance QoS [15], where a carriage is considered as a femtocell and signals inside the train are forwarded from the outside by cables.

*Doppler Frequency Offset and Spread*: 5G-R should support mobility up to 500 km/h. The corresponding Doppler frequency offset and spread affect demodulation performance of receiver in dynamic railway environments. For 2.1 GHz, the uplink Doppler frequency offset would be larger than 680 Hz for 350 km/h, and it would be much higher for millimeter-wave bands. The solution is mainly to use algorithms to adjust frequency of user and to eliminate the influence of Doppler offset.

*Multi-User Group Handover*: 5G-R base station spacing is reduced compared with GSM-R because of higher carrier frequency, and frequent group handover exists, which causes network congestion and performance degradation. Wireless network planning is required to achieve faster cell re-selection and reasonable cell overlap to satisfy inter-cell handover requirements. At the same time, cell merging can reduce the number of inter-cell handovers and improve performance and reliability. Moreover, new scheme are required to reduce the number of handover blockings, and femtocell-based network mobility scheme can be designed to support seamless handover in 5G-R scenarios.

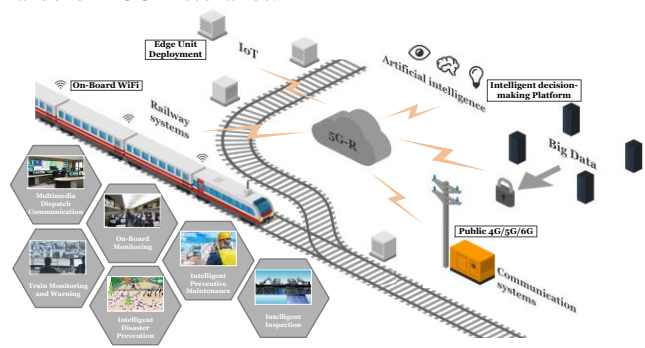

**Fig. 4.** Railway intelligent perception and decision-making platform.

*C. Intelligent Perception and Decision-Making Platform*

The future railway system needs to establish a complete intelligent perception and decision-making platform, to support full-element and full-status perception and decision-making of intelligent railway. A proposed vision of the platform is illustrated in Fig. 4. On this platform, railway systems, communication systems, IoT, big data and artificial intelligence technologies will be deeply integrated, to realize perception, transmission, and processing of different types, dimensions, and volumes of data. Artificial intelligence can enable the network to process large amounts of data, dynamically identify and adapt to complex scenarios, and satisfy the requirements for high-speed and real-time signal processing capabilities. Moreover, based on networking integrated cloud-edge-end, the intelligent perception and decision-making platform will support railway intelligent decision-making and signal control, and comprehensively improve intelligence levels of railway disaster prevention, operation and maintenance, monitoring, and dispatching. Such platform will motivate the evolution of 5G-R to 6G era with higher network intelligence.

*D. Integration of Advanced Technologies and Systems*

*Broadband and Narrowband Coexistence*: GSM-R and 5G-R networks will still have a long time to coexist and many issues need to be solved as follows: interconnection between GSM-R and 5G-R networks, cohabitation between GSM-R and 5G-R, overlapping coverage of GSM-R and 5G-R networks, collaboration of multimedia dispatch communications for GSM-R and 5G-R, updates of on-board terminal module to support both GSM-R and 5G-R, antenna updates and deployments for new frequency bands, etc. In addition, the future railway communications should be adaptable and can support multiple access technologies with bearer independence.

*Integration of Advanced Technologies*: It involves the integration and cooperation of 5G-R and different advanced technologies such as LTE, WiFi6, fixed 5G (F5G), satellite communications, millimeter-wave/terahertz communications, and IoT. 5G New Radio and LTE offer natural integration via the radio interface. WiFi6 achieves 9.6 Gbps data rate, which is an effective supplement for 5G New Radio networks in low-speed scenarios. It can be applied to 5G-R scenarios such as virtual reality, augmented reality, and video surveillance. F5G, based on passive optical network, is a fixed network supplement to 5G New Radio networks. With F5G, 5G-R can support massive connectivity and high capacity, and establishes railway Internet of Everything. Satellite communications such as low-orbit satellite Internet and high-throughput satellites can be integrated with 5G-R to develop the air-space-ground integrated railway communication network, which provides comprehensive and seamless coverage for different 5G-R scenarios. Moreover, a satellite navigation system combined with 5G-based positioning can realize high-precision positioning for various railway environments such as open areas and tunnels. Millimeter-wave/terahertz communications can realize ultra-high-speed transmission of on-board data and high-speed video data delivery of Passenger Information System. IoT can be used to realize the interconnection and perception of railway infrastructure, devices, equipment, etc.

## VI. CONCLUSION

This article has introduced 5G-R technology which could lead to fundamental changes in the further railway systems. 5G-R offers highly competitive performance, support many railway services, and can be applied for various railway application scenarios. In addition, intelligence will play a crucial role for 5G-R in increasing performance and improving services of the network. 5G-R system configurations are compared and discussed. Some key 5G-based technologies such as network architecture, massive MIMO, millimeter-wave, multiple access, ultra-reliable low latency communication, and video processing are discussed for railway applications, and a suite of these solutions will form the basis of 5G-R to satisfy the expected performance requirements. Finally, some technical challenges of 5G-R research and implementation are discussed.